\documentclass[aps, prx, reprint,footinbib,floatfix,superscriptaddress]{revtex4-2}
\usepackage{lipsum}
\usepackage{graphicx}
\usepackage{indentfirst}
\usepackage{braket}
\usepackage{float}
\usepackage{amsmath}
\usepackage{physics}
\usepackage{amssymb}
\usepackage{CJK}
\usepackage{esint}
\usepackage{color}
\usepackage[T1]{fontenc}
\usepackage{amsfonts}
\usepackage{footmisc}
\usepackage{scrextend}
\usepackage{multirow}
\usepackage[outdir=./]{epstopdf}
\usepackage{xcolor}
\usepackage[english]{babel}
\usepackage{url}
\usepackage{comment}
\usepackage{array}
\usepackage{subfiles}
\usepackage{tikz}
\usetikzlibrary{shapes,arrows}
\usepackage{mathrsfs}
\usepackage{mathtools}
\usepackage[mathscr]{eucal}
\usepackage{multirow}
 \usepackage{booktabs} 

\usepackage[hyperfootnotes=false]{hyperref}
\usepackage{cleveref}
\definecolor{darkblue}{rgb}{0,0,0.5}
\hypersetup{
colorlinks=true,
linkcolor=black,
filecolor=blue,
citecolor=darkblue,  
urlcolor=cyan,
}

\usepackage{trimclip}

\makeatletter
\DeclareRobustCommand{\shortto}{%
  \mathrel{\mathpalette\short@to\relax}%
}

\newcommand{\short@to}[2]{%
  \mkern2mu
  \clipbox{{.5\width} 0 0 0}{$\m@th#1\vphantom{+}{\shortrightarrow}$}%
  }
\makeatother

\usepackage{pict2e,picture,graphicx}

\makeatletter
\DeclareRobustCommand{\Arrow}[1][]{%
\check@mathfonts
\if\relax\detokenize{#1}\relax
\settowidth{\dimen@}{$\m@th\rightarrow$}%
\else
\setlength{\dimen@}{#1}%
\fi
\sbox\z@{\usefont{U}{lasy}{m}{n}\symbol{41}}%
\begin{picture}(\dimen@,\ht\z@)
\roundcap
\put(\dimexpr\dimen@-.7\wd\z@,0){\usebox\z@}
\put(0,\fontdimen22\textfont2){\line(1,0){\dimen@}}
\end{picture}%
}
\makeatother

\def\be{\begin{equation}}
\def\ee{\end{equation}}
\def\ba{\begin{eqnarray}}
\def\ea{\end{eqnarray}}
\def\bal{\begin{equation}\begin{aligned}}
\def\eal{\end{aligned}\end{equation}}

\def\bp{\begin{pmatrix}}
\def\ep{\end{pmatrix}}

\usepackage{bm}

\usepackage[normalem]{ulem}
\usepackage{xr}
\begin{document}

\title{
Resource-Efficient Quantum-Enhanced Compressive Imaging via Quantum–Classical co-Design
}

\author{
Haowei Shi$^{1}$ \quad
Visuttha Manthamkarn$^{2}$ \quad
Christopher M. Jones$^{3}$ \quad
Zheshen Zhang$^{2}$ \quad
Quntao Zhuang$^{1,4}$ \\
$^{1}$University of Southern California \quad
$^{2}$University of Michigan \\
$^{3}$Halliburton Technology \quad
$^{4}$University of Southern California\\
{\tt\small haow.shi@gmail.com, qzhuang@usc.edu}
}

\begin{abstract}

Quantum sensing can enhance imaging performance by reducing measurement noise below the classical limit, thereby improving the signal-to-noise ratio (SNR) of acquired data. In conventional quantum imaging schemes, squeezing is applied independently to each pixel or spatial mode, leading to a quantum resource cost that scales linearly with image dimension. This approach implicitly separates quantum enhancement from classical post-processing, treating them as independent layers. In this work, we demonstrate that integrating quantum resource allocation with the guidance from classical compressive imaging, via co-design between the quantum hardware layer and the classical software layer, substantially reduces the required quantum resources. We employ principal component analysis (PCA) to identify a low-dimensional principal component subspace for measurement and apply squeezing selectively to the most informative spatial modes corresponding to these principal components. Our numerical experiments show that high-accuracy image classification and high-fidelity image reconstruction can be achieved with significantly fewer squeezed modes compared to pixel-wise squeezing. Our results establish a joint quantum–classical co-design framework for resource-efficient quantum-enhanced imaging.
\end{abstract}

\maketitle

\section{Introduction}

Imaging performance is fundamentally constrained by measurement noise~\cite{cox1986information,tsang2016quantum}.
In photon-starving regimes, quantum fluctuations impose a shot-noise floor that cannot be removed by classical post-processing algorithms or system engineering alone.
As a result, improving the signal-to-noise ratio (SNR) at the sensing stage is a central challenge across a wide range of imaging applications, including biological microscopy, scientific instrumentation, and remote sensing.

Quantum sensing offers a promising route to surpass classical noise limits by exploiting nonclassical states of light.
By reducing measurement uncertainty below the standard quantum limit, quantum resources can enhance the quality of acquired measurements and consequently improve the performance of downstream imaging and vision tasks~\cite{shi2025quantum}.
Higher-SNR measurements can translate into improved reconstruction accuracy and more reliable inference in data-driven imaging pipelines.

Modern computational imaging systems increasingly integrate optical sensing with machine learning-based reconstruction~\cite{sitzmann2018end,chakrabarti2016learning}.
Rather than treating sensing hardware and image processing algorithms as independent components, recent work in computational imaging highlights the importance of jointly designing sensing and reconstruction pipelines.
However, most existing computational imaging frameworks operate under classical sensing limits.
In photon-starved regimes---such as low-light microscopy or long-range remote sensing---quantum sensing provides a potential mechanism to reduce measurement noise below the standard quantum limit.
This raises a new design question: how should limited quantum resources be allocated when the downstream objective is vision-based inference rather than full image acquisition?

A common approach to quantum-enhanced imaging is to apply quantum noise reduction independently across all spatial modes or pixels, thereby uniformly lowering the measurement noise floor.
In optical implementations, this is often achieved using squeezed light, which reduces noise in a chosen quadrature~\cite{Caves1981QuantumNoise,collett1987quantum,weedbrook2012gaussian}.
Squeezed-vacuum injection has enabled quantum-enhanced precision measurements in large-scale gravitational-wave interferometers~\cite{aasi2013enhanced,Tse2019QuantumEnhancedLIGO,Acernese2019VirgoReach} and spectroscopy~\cite{yang2021squeezed,hariri2025entangled}, with up to 15\,dB squeezing demonstrated experimentally~\cite{Vahlbruch2016Detection15dB}.

However, straightforward pixel-wise squeezing leads to a major scalability bottleneck: the number of squeezed modes required grows linearly with image dimension.
Such resource demands quickly become prohibitive in practical high-dimensional imaging systems.

Conceptually, quantum-enhanced imaging is often viewed as a two-layer architecture.
The first layer---the \emph{quantum layer}---improves the SNR of measurements through nonclassical resources.
The second layer---the \emph{classical layer}---reconstructs the image using signal processing or learning-based algorithms.
Under this viewpoint, the quantum layer typically executes identical quantum state preparation operations for each pixel independently, assuming no prior knowledge about how the measurements will be used during reconstruction.

Meanwhile, classical compressive imaging provides a scalable strategy by exploiting low-dimensional structure in natural images.
Many images admit sparse or approximately low-rank representations, enabling accurate recovery from a small number of informative projections~\cite{Donoho2006CS,Candes2006RobustCS,Candes2006NearOptimal}.
This observation suggests that not all spatial modes contribute equally to downstream imaging tasks.

In this work, we bridge the conventional separation between the quantum sensing layer and the classical reconstruction layer.
We show that image statistics learned from data can directly guide how quantum resources are allocated.
By co-designing quantum enhancement with data-driven compressive reconstruction, we significantly reduce the quantum resources required for imaging.

Specifically, we employ principal component analysis (PCA) to learn a low-dimensional measurement subspace from training data and apply squeezing selectively only to the first $k$ principal-component modes, where $k$ is much smaller than the number of pixels $d$.
This strategy compresses the measurement stage itself, reducing the number of homodyne readouts and squeezed sources while preserving reconstruction fidelity.
Unlike classical compressive imaging, the dimensionality reduction here is implemented physically in the quantum sensing layer through beamsplitter transformations and mode-selective squeezing, enabling direct reduction of quantum hardware resources rather than only computational complexity.
Our results demonstrate that high-quality quantum-enhanced image reconstruction can be achieved with substantially fewer squeezed modes compared to conventional pixel-wise squeezing approaches.
This joint quantum--classical framework provides a scalable path toward resource-efficient quantum-enhanced compressive imaging.

\textbf{Contributions.} Our main contributions are:

\begin{itemize}
\item We introduce a quantum–classical co-design framework that integrates quantum resource allocation with data-driven compressive imaging.
\item We propose a PCA-guided quantum sensing architecture that selectively allocates squeezing to informative principal-component modes, reducing the required number of squeezed sources from $d$ to $k \ll d$.
\item Through experiments on MNIST and CIFAR-10 tasks, we demonstrate that the proposed method preserves downstream classification and reconstruction performance while substantially reducing quantum resource requirements.
\end{itemize}

This paper is organized as follows.
In \Cref{sec:relatedwork}, we review prior research on sensor--algorithm co-design and quantum-enhanced pattern recognition, and position our contribution relative to these lines of work.
In \Cref{sec:task}, we formulate the imaging task.
In \Cref{sec:preliminary}, we introduce the preliminaries of principal component analysis and summarize the quantum sensing model underlying our framework.
In \Cref{sec:theory}, we present our quantum--classical co-design methodology, including PCA-guided mode selection, optical implementation details, and resource-scaling analysis.
In \Cref{sec:experiment}, we report numerical results on MNIST and CIFAR-10 for classification and denoising, comparing quantum-enhanced and classical baselines under matched settings.
Finally, \Cref{sec:conclusion} concludes the paper.

\section{Related Works}
\label{sec:relatedwork}

\textbf{Sensor-algorithm co-design:}
Recent works in computational imaging have demonstrated that sensing and reconstruction should not be designed independently. Instead of treating the optical front-end as fixed and optimizing only the post-processing algorithm, these works jointly optimize the physical measurement process and the reconstruction pipeline to improve overall system performance. 

For example, Sitzmann et al.~\cite{sitzmann2018end} propose an end-to-end framework that jointly optimizes diffractive optical elements and image reconstruction networks using differentiable optical simulations. By co-designing optics and computation, they achieve improved extended depth-of-field and super-resolution performance compared to sequential design approaches. Similarly, Chakrabarti~\cite{chakrabarti2016learning} learns sensor multiplexing patterns through back-propagation, jointly optimizing the sensor layout and reconstruction network to surpass conventional hand-designed color filter arrays such as the Bayer pattern. 

These works illustrate a broader principle: measurement design should be task- and data-aware, rather than fixed a priori. Our work adopts a similar system-level perspective, but in the context of quantum-enhanced imaging. Instead of jointly optimizing classical optics and reconstruction, we integrate quantum resource allocation with data-driven compressive reconstruction. In particular, we show that squeezing need not be applied uniformly across all spatial modes; rather, it can be selectively allocated to informative measurement subspaces identified from data, thereby reducing quantum resource consumption while preserving reconstruction fidelity.

\textbf{Entanglement-assisted pattern recognition:} In practical settings, the key benefit of quantum-enhanced imaging comes from reducing the effective measurement noise floor (e.g., shot noise) through squeezing~\cite{Caves1981QuantumNoise,Marino2019QuantumSensing}. This can increase the available signal-to-noise ratio (SNR) for a fixed photon budget, which is particularly relevant when illumination power is constrained by sample damage or safety considerations.

Ref.~\cite{banchi2020quantum} provides an instructive example of bridging the SNR advantage of quantum optical resources and the task of reading classical information patterns (e.g., barcodes) under photon-number constraints, showing that quantum illumination using two-mode squeezed states can improve decoding/discrimination performance compared with classical strategies. This line of work highlights that quantum resources can be most impactful when the task depends on a limited set of informative spatial features rather than full image reconstruction. 

Ref.~\cite{ortolano2023quantum} studies quantum advantage in pattern recognition. In their framework, entanglement (via two-mode squeezed states) is used to generate correlated measurement outcomes that can improve discrimination among a set of hypotheses compared with classical illumination under comparable constraints. While their goal is pattern recognition rather than full image reconstruction, the underlying message is aligned with quantum-enhanced imaging: quantum resources can yield substantial advantages in pattern recognition accuracy due to the gain in SNR.

\textbf{Our approach:} Our paper extracts the similar SNR advantage from quantum sources, but targets a different bottleneck: the scalability of applying squeezing across high-dimensional pixel spaces. Instead of attempting to squeeze each pixel mode (which scales with the number of pixels), we use PCA to identify a low-dimensional subspace that captures most of the task-relevant image variance. We then allocate squeezed sources only to this reduced set of principal-component modes, aiming to preserve the quantum SNR advantage while reducing the required number of squeezed resources and the optical complexity. Furthermore, our ultimate task is the full image reconstruction, beyond the much simpler classification task of pattern recognition.

\section{Task Formulation}
\label{sec:task}
We consider an imaging system that measures an object through a quantum-enhanced sensing pipeline. 
Let $x \in \mathbb{R}^d$ denote a $d$-pixel image represented as a vector of spatial measurements.

In conventional optical sensing, homodyne measurements are corrupted by quantum shot noise.
Under vacuum-state probing, each pixel of image has inherent noise, with unit normalized to
$
\mathrm{Var}(X) = 1/2.
$
Quantum squeezing of gain $G$ allows the noise reduction in each pixel measurement to
$
\mathrm{Var}(X) = {1}/{2G},
$
which improves the effective signal-to-noise ratio of the acquired measurements.

In conventional quantum-enhanced imaging, squeezing is applied pixel-wise.
For a $d$-pixel image, this requires $d$ squeezed-state sources and $d$ homodyne detectors.
Such requirement is impractical for high-dimensional imaging systems.

Our goal is to design a sensing strategy that maximizes downstream imaging performance while using only a limited number of squeezed sources.
Formally, given a sensing system with at most $k$ squeezed modes where $k \ll d$, we seek an entangled probe design and measurement strategy that preserves the task-relevant information of the image while minimizing the required quantum resources.

\section{Preliminaries}
\label{sec:preliminary}

\subsection{Principal Component Analysis}

Principal component analysis (PCA) is a classical dimensionality-reduction method that dates back to Pearson's original formulation~\cite{Pearson1901PCA} and was further developed by Hotelling~\cite{Hotelling1933PCA}. Given a centered data matrix $\bm X\in\mathbb{R}^{N\times d}$ (e.g., $N$ training images each reshaped into a $d$-pixel vector), PCA finds an orthonormal basis that diagonalizes the sample covariance $\bm \Sigma=\frac{1}{N}\bm X^{\top}\bm X$. Writing the eigendecomposition $\bm \Sigma=\bm U \bm \Lambda \bm U^{\top}$ with eigenvalues $\lambda_1\ge\lambda_2\ge\cdots\ge\lambda_d$, the $j$th principal component corresponds to the eigenvector $\bm u_j$; projecting an image vector $\bm x\in\mathbb{R}^d$ onto this basis yields coefficients
$
\bm z=\bm U^{\top} \bm x,
$
k-PCA retains only the first $k$ components and gives the rank-$k$ approximation
$
\tilde {\bm x}= \sum_{j=1}^k z_j \bm u_j ,
$
which is optimal (in mean-squared error) among all rank-$k$ linear approximations by the Eckart--Young theorem~\cite{EckartYoung1936SVD}. A standard interpretation is that the ratio
$\sum_{j=1}^k\lambda_j/\sum_{j=1}^d\lambda_j$ quantifies the fraction of total variance explained by the first $k$ components~\cite{Jolliffe2002PCA}.

\begin{figure}[!htbp]
    \centering
    \includegraphics[width=\linewidth]{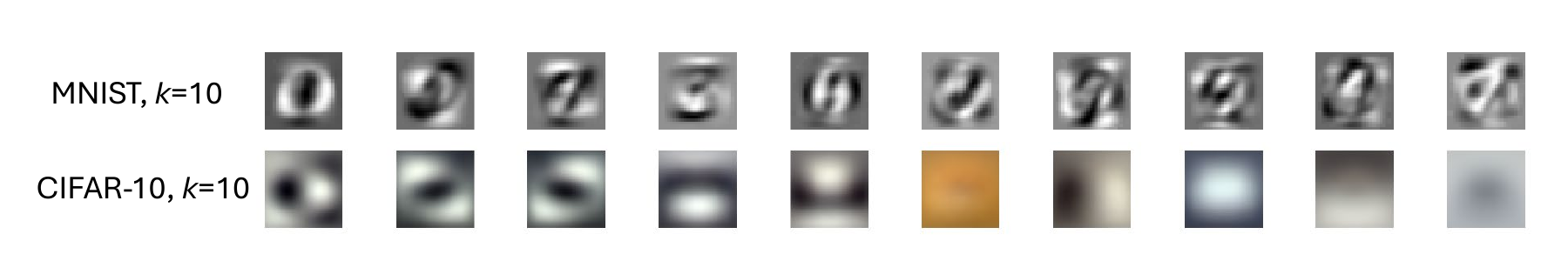}
    \caption{The first 10 principal components of MNIST and CIFAR-10 datasets.}
    \label{fig:PCs}
\end{figure}

For the MNIST dataset and CIFAR-10 dataset used in this paper, we illustrate the first 10 principal components in Fig.~\ref{fig:PCs}. The PCA-inferred quantum layer is to be conducted over the whole dataset including both the training set and the test set, while the PCA eigenvectors are calculated over only the training set.

\subsection{Quantum Squeezing}
In quantum optics, following the continuous-variable formalism in Ref.~\cite{weedbrook2012gaussian}, an electromagnetic field is quantized into a single bosonic mode, described by the annihilation operator $\hat a$ and the (dimensionless) field quadratures
\begin{equation}
\hat q=\sqrt{2}\Re \hat a=\frac{\hat a+\hat a^{\dagger}}{\sqrt{2}},\qquad
\hat p=\sqrt{2}\Im \hat a=\frac{\hat a-\hat a^{\dagger}}{\sqrt{2}i},
\end{equation}
Physically, the quadratures $\hat q,\hat p$ are proportional to the electric field $E$ and the magnetic field $H$ respectively, which satisfy $[\hat q,\hat p]=i$. For the vacuum state, the quadrature means are zero, while the quadrature variances are shot-noise-limited,
\begin{equation}
\mathrm{Var}(\hat q)=\mathrm{Var}(\hat p)=\frac{1}{2}.
\end{equation}
Its quasi-probability distribution forms a Gaussian distributed disk at the center of the phase space spanned by $\hat q,\hat p$.
A (single-mode) squeezed state is generated by the squeezing operator $\hat S(r)=\exp\!\left[\frac{r}{2}(\hat a^2-\hat a^{\dagger 2})\right]$, which redistributes quantum fluctuations between conjugate quadratures while preserving the Heisenberg uncertainty relation. For a state squeezed along $\hat q$ (squeezing parameter $r>0$), one has
\begin{equation}
\mathrm{Var}(\hat q)=\frac{1}{2}e^{-2r}=\frac{1}{2G},\qquad
\mathrm{Var}(\hat p)=\frac{1}{2}e^{2r}=\frac{G}{2},
\end{equation}
so the noise in $\hat q$ is reduced below the vacuum level at the expense of increased noise in $\hat p$, where we define the squeezing gain $G=e^{2r}\ge 1$. The quasi-probability distribution of a squeezed state forms a squeezed disk, an ellipse, at the center of the phase space.

\subsection{Entanglement }
\begin{figure}[!htbp]
    \centering
    \includegraphics[width=\linewidth]{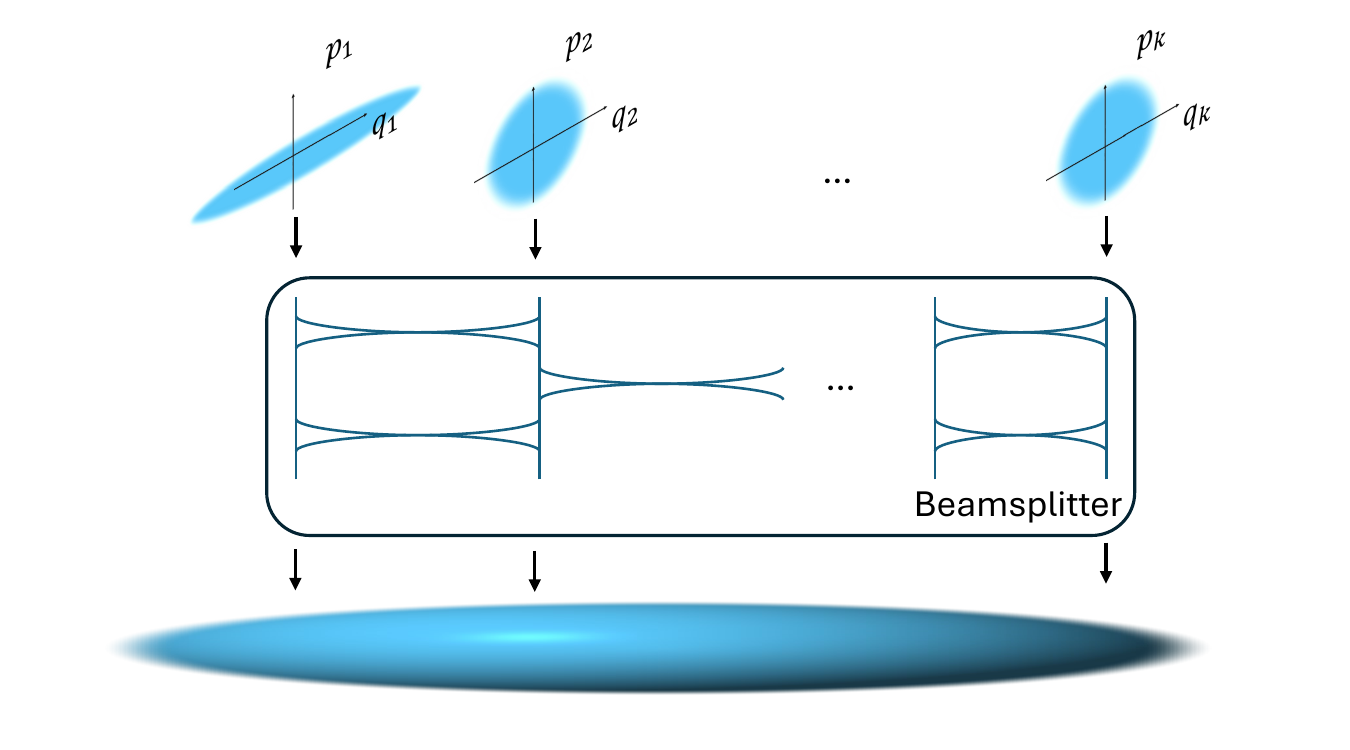}
    \caption{Mixing a mode in squeezed state with modes in vacuum state generates entanglement. The EPR quadrature of the output is squeezed.}
    \label{fig:entanglement}
\end{figure}

A simple example of continuous-variable entanglement is the two-mode squeezed state, which is also known as the continuous-variable EPR state since the EPR quadratures of it are squeezed~\cite{weedbrook2012gaussian}
\bal 
{\rm Var} (\hat q_-) =\frac{1}{2G}\,,\quad {\rm Var} (\hat p_+) =\frac{1}{2G}
\eal 
where $\hat q_-=\frac{\hat q_1-\hat q_2}{\sqrt{2}}$, $\hat p_+=\frac{\hat p_1+\hat p_2}{\sqrt{2}}$.

In our protocol, we use a converging beamsplitter to implement the projection to the principal subspace of the PCA, which outputs the multi-mode EPR quadratures, a linear combination of the local quadratures of input as shown later in Eq.~\eqref{eq:bs}. The final heterodyne detection measures these EPR quadratures.
The squeezing on the EPR quadratures requires a global entanglement. This can be generated by a diverging beamsplitter, as shown in Fig.~\ref{fig:entanglement}.
For each specific dataset, we will do the data-driven PCA and tune the global entanglement accordingly to minimize the noises on the EPR quadratures to be measured.

\subsection{Squeezing-enhanced RF-photonics}

\begin{figure*}[!htbp]
    \centering
    \includegraphics[width=\linewidth]{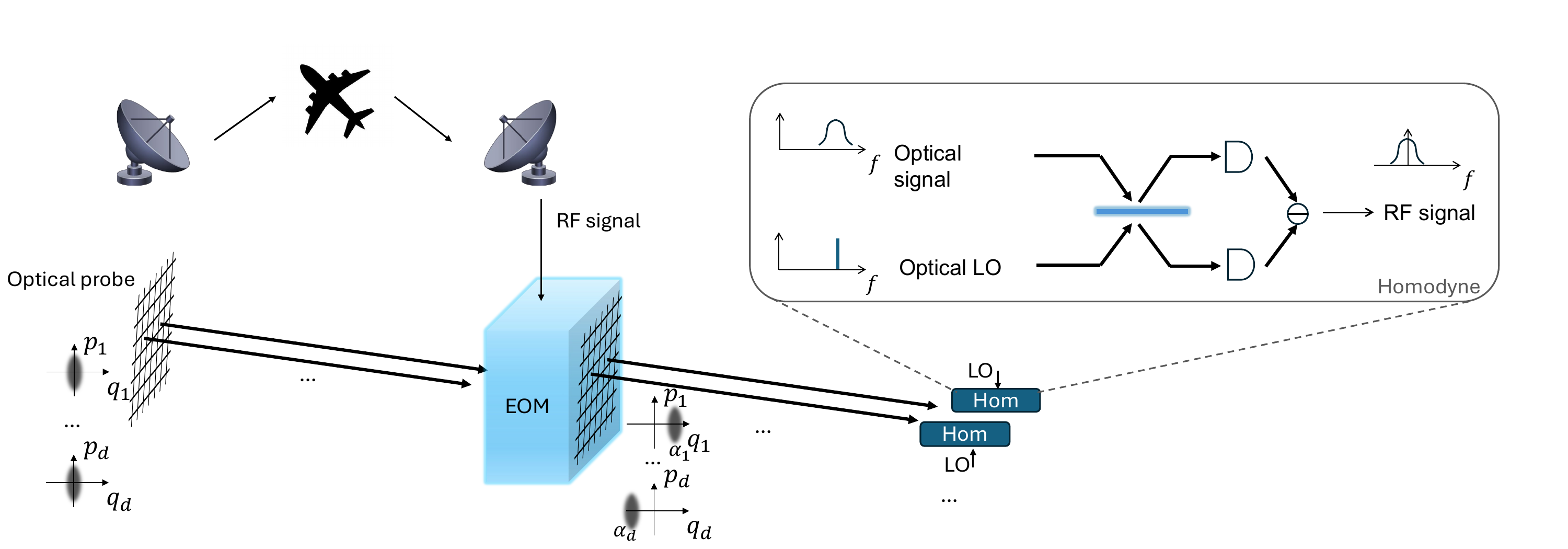}
    \caption{Schematic of conventional pixel-wise squeezing-enhanced RF-photonics with a pixel-wise homodyne detector array. EOM: electro-optic modulator. LO: local oscillator.}
    \label{fig:homodyne}
\end{figure*}

Beyond optical imaging, quantum resources have also been explored in radio-frequency (RF) photonics, where optical carriers enable coherent sensing, distribution, and readout of RF signals. At optical frequencies the thermal photon number at room temperature is negligible, so photonic sensing is limited by quantum vacuum noise (shot noise) rather than thermal noise. Because RF-to-optical transduction can be inefficient, this shot-noise floor often dominates the receiver noise. Coherent measurements that access both optical quadratures are therefore required to recover RF information encoded in the optical amplitude and phase. Ref.~\cite{xia2020demonstration} demonstrated a reconfigurable entangled RF photonic sensor network that distributes nonclassical correlations across sensor nodes to enhance sensitivity. Building on this idea, Ref.~\cite{shi2025quantum} proposed quantum-enhanced RF photonic distributed imaging, showing that squeezing or entanglement can reduce measurement noise and improve imaging performance under photon-number constraints.

The conventional pixel-wise squeezing-enhanced RF photonic system is illustrated in Fig.~\ref{fig:homodyne}. An RF probe signal interrogates the object and the returned field $\hat b(t;x,y)$ is transduced to the optical domain by an electro-optic modulator (EOM). The resulting optical signal is
$
\hat a(t;x,y)=\hat a_{\rm RF}(t;x,y)e^{-i(\omega_c t+\theta)},
$
whose mean value encodes the RF information $\expval{\hat a_{\rm RF}(t;x,y)}\propto \expval{\hat b(t;x,y)}$. For a discrete image of $d$ pixels we discretize $(x,y)$ into $\{(x_j,y_j)\}_{j=1}^d$. Balanced homodyne detection provides a quadrature-selective measurement using a strong local oscillator $B(t)=Be^{-i\omega_c t}$, yielding the photocurrent
$
\hat I(t;x,y)=|\hat a(t;x,y)+B(t)|^2-|\hat a(t;x,y)-B(t)|^2
=B(\hat a_{\rm RF}(t;x,y)e^{-i\theta}+\hat a^\dagger_{\rm RF}(t;x,y)e^{i\theta})
\propto \hat q_{\rm RF}(t;x,y)\cos\theta+\hat p_{\rm RF}(t;x,y)\sin\theta,
$
where the real and imaginary quadratures are $\hat q= \sqrt{2}\Re\hat a=\frac{1}{\sqrt2}(\hat a+\hat a^\dagger)$ and $\hat p=\sqrt{2}\Im\hat a=\frac{1}{\sqrt2 i}(\hat a-\hat a^\dagger)$. The relative phase $\theta$ selects the measured quadrature $\hat X_\theta=\hat q\cos\theta+\hat p\sin\theta$. Since time encoding is not considered here, the time variable $t$ is omitted in the following.

In the shot-noise-limited regime the measurement variance is $\mathrm{Var}(\hat X_\theta)=\frac12$. If the probe is prepared in a squeezed state with gain $G$ aligned with the measured quadrature, the variance becomes
\begin{equation}
\mathrm{Var}(\hat X_\theta)=\frac{1}{2G}<\frac12,
\end{equation}
reducing measurement noise and improving the effective SNR for estimating the image-dependent quadrature displacement.

However, conventional pixel-wise squeezing requires $d$ squeezed states, so the resource cost scales linearly with image dimension and becomes impractical for high-resolution imaging. In this work we address this limitation using PCA-based image dimension reduction.

\section{Methods}
\label{sec:theory}
\subsection{Quantum-Enhanced Compressive Imaging Based on PCA}

\begin{figure*}[!htbp]
    \centering
    \includegraphics[width=\linewidth]{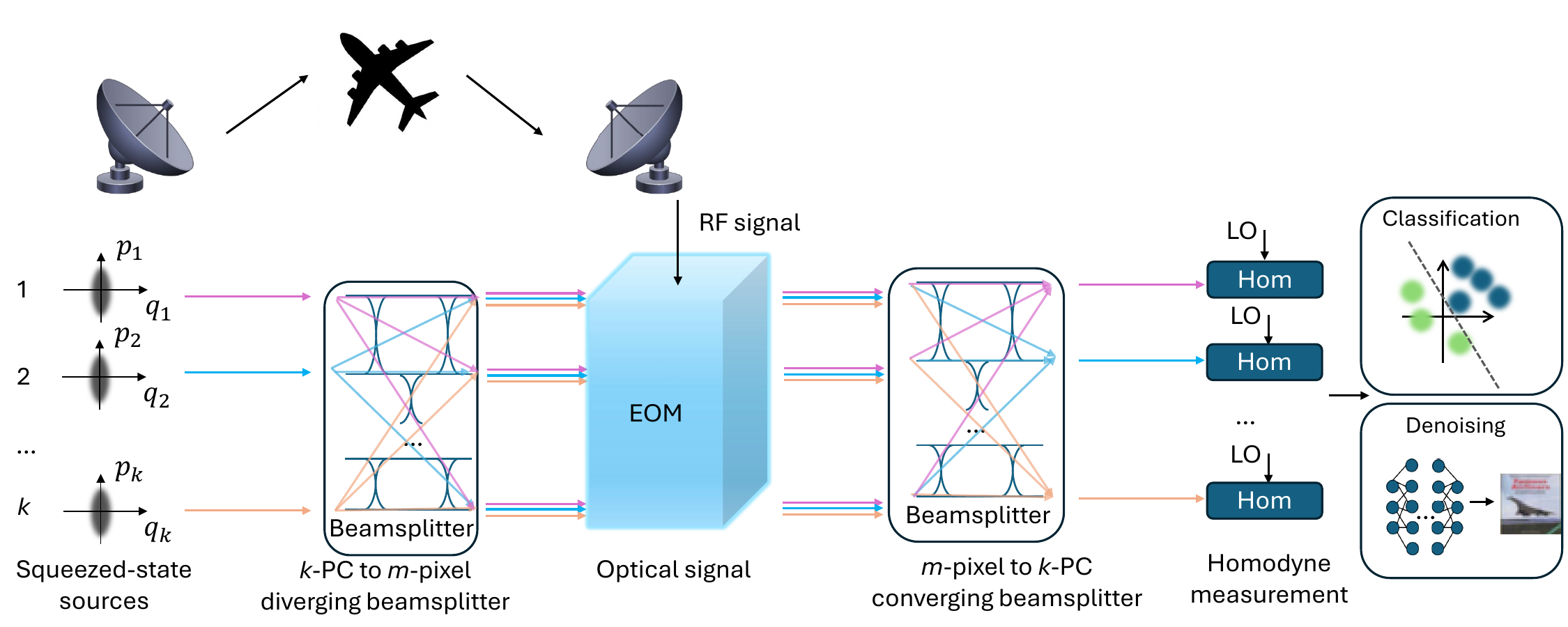}
    \caption{Schematic of the PCA imaging. EOM: electro-optic modulator. LO: local oscillator.}
    \label{fig:schematic_PCA}
\end{figure*}

The schematic of the quantum-enhanced compressive imaging is shown in Fig.~\ref{fig:schematic_PCA}. 
First, a transmitter antenna sends an RF signal to the object and then a receiver antenna obtains the returned RF signal $\hat b(x,y)$ that carries the information about the object. Meanwhile, considering $k$ principal components, we prepare $k$ squeezed-state point sources as the probes in the optical domain, with $\text{Var}\Re\hat a_{{\rm in},j}=1/4G$, then we use the diverging beamsplitter to distribute them to the $k$ spatial modes corresponding to the $k$ principal components solved from PCA. Formally, the beamsplitter fulfills the symplectic transform on the field annihilation operators 
\be 
\{\hat a_{{\rm in}, j}\}_{j=1}^k\to \hat a_{\rm probe}(x,y)= \sum_{j=1}^k f_j(x,y) \hat a_{{\rm in}, j} 
\label{eq:bs}
\ee 
where $f_j(x,y)$ is the spatial distribution of the $j$th principal component (PC), which satisfies the unitary condition $\int dx dy |f_j(x,y)|^2=1$ for all $j$. In the middle, we modulate the RF signal on the distributed quantum probes using an EOM, so that the information is transduced from the RF domain to the optical domain
\be 
\hat a_{\rm probe}(x,y)\to \hat a_{\rm probe}'(x,y)= \hat a_{\rm probe}(x,y) + \alpha(x,y)
\ee 
where $\alpha(x,y)\propto \expval{\hat b(x,y)}$ is the complex field displacement induced by the RF signal. In this work we assume perfect phase locking, so that $\alpha(x,y)$ are real numbers corresponding to the image data.
Finally, we use a converging beamsplitter to convert the returned probe field back to $k$ points 
\bal  
~&\hat a_{\rm probe}'(x,y)\to \\
&\left\{\hat a_{{\rm out},j}|\hat a_{{\rm out},j}= \int dx dy f_j^*(x,y)  \hat a_{\rm probe}'(x,y)\right\}_{j=1}^k\,,
\eal
and then make homodyne measurements on them. The mean values of the $k$ homodyne measurements give the $k$ principal components of the RF signal $\alpha(x,y)$
\be 
\expval{\hat a_{{\rm out},j}} =  \int dx dy f_j(x,y)  \alpha(x,y)\equiv z_j\,.
\ee 
Since the converging and diverging beamsplitters are inverse to each other, the variances are recovered to the input squeezed variances $\text{Var}~\hat a_{{\rm out},j} = \text{Var}~\hat a_{{\rm in},j} $.
After obtaining the measured principal components, one can implement image classification or denoising via supervised learning, e.g. support vector machine (SVM) and neural network. In Sec.~\ref{sec:experiment}, we will evaluate the quantum enhancement for them.

\paragraph{Reduction of the required number of measurements using PCA.---}
If the image ensemble is well-approximated by a $k$-dimensional principal subspace (i.e., the information needed for the task is sufficiently captured by $\{ f_1(x,y), \ldots, f_k(x,y)\}$), then the task-relevant degrees of freedom are the $k$ coefficients $\{z_j\}$ rather than all $d$ pixel values. In our optical implementation, homodyne detection after the beamsplitter that maps the spatial field into the principal component basis $f_j(x,y)$ effectively measures these coefficients; thus, one can reduce the number of homodyne measurements from $d$ (pixel-wise probing) to $k\ll d$ (PC-wise probing) while still enabling sufficiently accurate reconstruction via inverse PCA operation: $\tilde \alpha(x,y)=\sum_{j=1}^k \bm f_j(x,y) z_j$.

\paragraph{Cropping the principal component vector.---} Consider $d$-pixel images. In a vanilla PCA, each principal component vector is $d$-dimensional, representing a pattern over the full $d$-pixel space. However, in quantum information processing, the conversion from the image space to the principal component space is implemented by a beamsplitter, and it is challenging to require the beamsplitter to possess $d\gg 1$ ports, since the quantum interference between $d$ ports requires extremely high coherence and low loss.
Hence, we consider beamsplitter with $m$ pixels each in this paper, i.e. the principal component vectors are cropped from $d$-dimensional to $m$-dimensional. We define $m$-pixel $k$-PCA, such that the beamsplitter only covers the first $m$ pixels with largest weights for each of the $k$ principal components. 

\section{Experiments}
\label{sec:experiment}

\subsection{Setup of image dataset and noise}
We use the 10-class MNIST dataset for both image classification and image denoising, and the 10-class CIFAR-10 dataset for image classification. In homodyne measurement, the measured principal component signals are corrupted by a Gaussian noise, modeled using the quantum model. Such noise is squeezed to $1/2G$ given the squeezed-state source with squeezing gain $G$. To constrain the image signal physically, we define the signal photon number per image 
\be 
n_S\equiv \sum_{j=1}^d |\alpha(x_j,y_j)|^2
\ee 
where we set the signal $\alpha(x_j,y_j)$ proportional to the $d$-pixel image data at the $j$th pixel, for $1\le j\le d$.
We compare a quantum-enhanced setting that employs squeezed-state sources of $G=10$ against a classical benchmark that employs vacuum-state sources $G=1$. For the classification task, we also provide the accuracy of the noiseless case as reference, which sets the ultimate limit of the accuracy corresponding to infinite squeezing $G\to \infty$.

To avoid overfitting, we train the neural network on the training dataset with 50000 items, and evaluate the classification accuracy benchmark on the test dataset with 10000 items different from the training dataset. For MNIST dataset with 60000 training data and 10000 test data, we use the remaining 10000 items in the MNIST dataset as validation dataset to tune the hyperparameters. For CIFAR-10 dataset with 50000 training data and 10000 test data, we use up all the training data for training without validation, while the hyperparameters are optimized using the package PyTorch with function ``torch.optim.lr\_scheduler.OneCycleLR''.

\subsection{Image Classification of MNIST} 

To begin with, we present the result of image classification over the 10-class MNIST data as shown in Fig.~\ref{fig:PCA_accu}, under signal photon number per image $n_S=10$. The classification is implemented by applying a SVM on the homodyne measured principal components. 

\subsubsection{Implementation details}
We use the SVMs to implement the classification. SVMs are supervised classifiers that learn a decision boundary with maximum geometric margin. For binary-labeled training data $\{(\bm x_i,y_i)\}_{i=1}^N$ with $y_i\in\{\pm 1\}$, a linear SVM seeks a separating hyperplane $f(\bm x)=\bm w^{\top}\bm x+b$ by solving the soft-margin optimization
$
\min_{\bm w,b,\{\xi_i\}}\ \frac{1}{2}\|\bm w\|_2^2 + C\sum_{i=1}^N \xi_i\,,
$
$\text{s.t.}\quad
y_i(\bm w^{\top}\bm x_i+b)\ge 1-\xi_i,\ \xi_i\ge 0,
$
where $C>0$ controls the tradeoff between margin size and classification errors (slack variables $\xi_i$). 
To separate linearly inseparable data, the kernel trick replaces inner products $\bm w^{\top}\bm x_j$ with a positive-definite kernel $K(\bm w,\bm x_j)$, which corresponds to a linear separator in an implicit feature space. In our experiments, the SVM takes the homodyne measurement outcomes as the feature vector for classification, we choose the radial basis function (RBF) kernel
$
K(\bm w,\bm x)=\exp\left(-\gamma\,\|\bm w-\bm x\|_2^2\right),
$
where $\gamma=1 / (n_{\rm feature}  {\rm Var}(\bm X)) $ sets the kernel width, $\bm X$ is the dataset, $n_{\rm feature} =k$ is the feature vector size i.e. the number of principal components here.

\subsubsection{Experiment results}

\begin{figure}[!htbp]
    \centering
    \includegraphics[width=\linewidth]{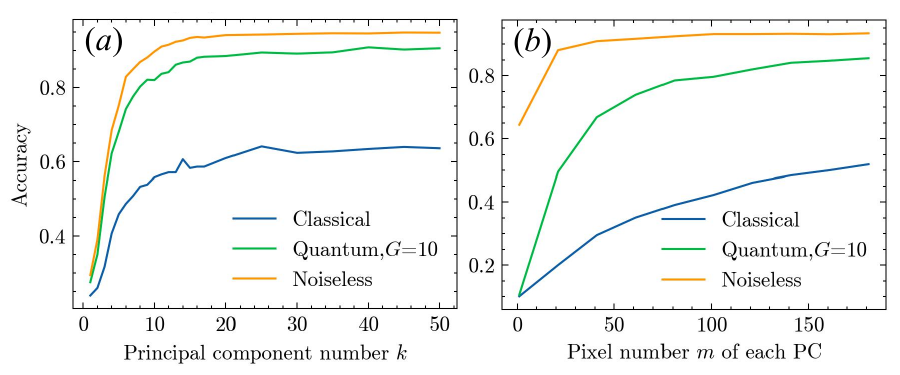}
    \caption{(a) Accuracy of the MNIST data classification with $k$-PCA imaging versus the principal component number $k$. Pixel number $m=d$ covers the whole image. (b) Accuracy of the MNIST data classification with $m$-pixel $15$-PCA imaging versus the beamsplitter pixel number $m$. Signal photon number $n_S=10$ per image.}
    \label{fig:PCA_accu}
\end{figure}

We show the accuracy curve of the MNIST dataset classification task in Fig.~\ref{fig:PCA_accu}. In subplot (a), we plot the accuracy versus the principal component number $k$. Compared with the classical benchmark with vacuum noise (blue), the quantum-enhanced case (green) with squeezed vacuum noise at merely an intermediate squeezing gain $G=10$ shows significant advantage in the accuracy rate, which is close to the ultimate noiseless limit (orange). As for the trend versus $k$, all three curves saturate at $k\simeq 20$, where the quantum-enhanced accuracy with $G=10$ is $\simeq 90\%$, classical accuracy $\simeq 60\%$. In subplot (b), we plot the accuracy versus the beamsplitter pixel number $m$ given the principal component number $k=15$, using same coloring for classical (blue), quantum (green), and noiseless (orange). As $m$ grows, we see that the quantum advantage becomes significant. The quantum-enhanced accuracy is saturated at $m\simeq 100$ to be $\simeq 80\%$. Note that the total pixel number here is $d=28\times 28=784$, our results show that the quantum advantage can be achieved with the resource cost of the quantum-squeezed source number $k\simeq 15$ and beamsplitter complexity $m\simeq 100$, significantly reduced from the conventional pixel-wise cost $d=784$.

\subsection{Image Classification of CIFAR-10}

\subsubsection{Implementation details}
We use a convolutional neural network to implement end-to-end recovery from the measured principal component signals to the label of the object. We adopt a prescribed neural network design, residual network (ResNet) with 18 layers, equipped with dropout layers to enhance generalizability, which is known to produce >90\% accuracy for the conventional CIFAR-10 datasets. To make the principal component data compatible with the input layer of the ResNet-18 which requires $3\times 32\times 32$ image input, we first process the principal components via inverse PCA to produce a raw estimation of image, then feed the raw image estimations towards the ResNet-18.

We use the PyTorch package to implement the neural network. Specifically, we import the resnet18 structure from torchvision.models, and revise its output layer from 100-output to 10-output to fit the 10 classes of our CIFAR-10 dataset. Finally, we add dropout layers after each activation with dropout rate 1\%, to suppress overfitting.

\begin{figure}[!htbp]
    \centering
    \includegraphics[width=.7\linewidth]{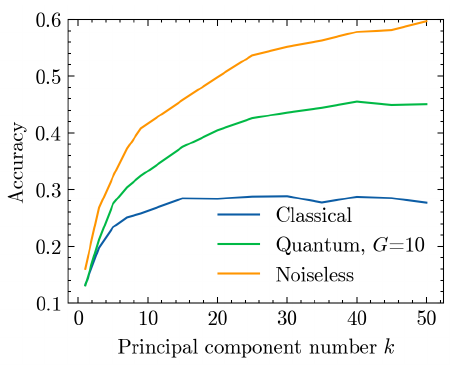}
    \caption{Accuracy of the CIFAR-10 data classification with $k$-PCA imaging versus the principal component number $k$. Signal photon number $n_S=10$ per image.}
    \label{fig:PCA_accu_CIFAR10}
\end{figure}

\subsubsection{Experiment results}
We show the accuracy curve of the CIFAR-10 dataset classification task in Fig.~\ref{fig:PCA_accu_CIFAR10}. Again, we observe significant quantum advantage with squeezing $G=10$, which is significantly below the noiseless limit, thus there is still a remarkable advantage margin that can be achieved by further increasing the squeezing gain $G$. At $k=50$, the quantum-enhanced accuracy with $G=10$ is $ 45\%$ while the classical accuracy is $28\%$. Note that the total pixel number here is $d=3\times 32\times 32=3072$, our results show that the quantum advantage can be achieved with the resource cost of the quantum-squeezed source number $k\simeq 50$ significantly reduced from the conventional pixel-wise cost $d=3072$.

\subsection{Image Denoising}
\begin{figure*}[!htbp]
    \centering
    \includegraphics[width=0.95\linewidth]{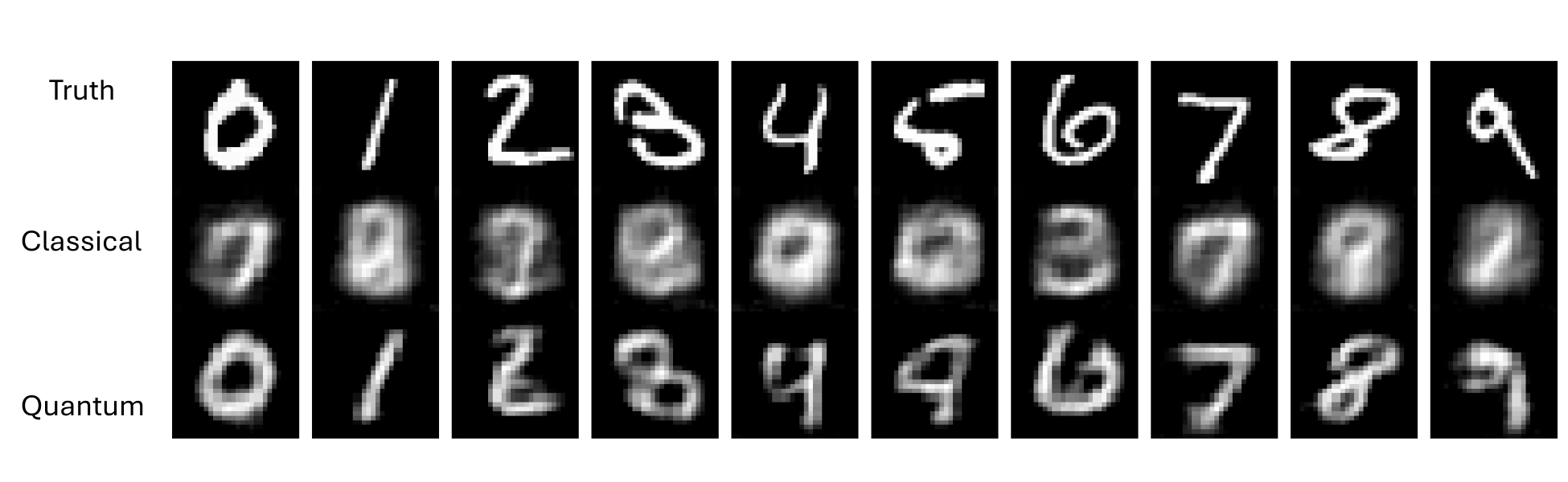}
    \caption{Imaging performance of the 50-PCA with convolutional neural network data processing. Signal photon number per image $n_S=1$.}
    \label{fig:PCA_recover_vs_m_n=1}
\end{figure*}

Now we present the result of image denoising over the 10-class MNIST data as shown in Fig.~\ref{fig:PCA_recover_vs_m_n=1}, under $n_S=1$ which gives lower SNR and demonstrates the best contrast between quantum-enhanced performance and the classical baseline. The denoising is implemented by applying a 4-layer convolutional neural network on the homodyne measured principal components given the ground truth as labels.

\subsubsection{Implementation details}

\begin{table}[!h]
\centering
\caption{Neural network architecture for denoising (decoder) with $k$-PCA input size $k=50$. Output is an image of size $28\times 28$.}
\label{tab:nn_arch_kpca50_decoder}
\begin{tabular}{l l l r}
\hline
\textbf{Layer (type)} & \textbf{Output shape} & \textbf{Param \#} \\
\hline
Linear-1 & $[-1,1,128]$ & 6,528 \\
Upsample-2 & $[-1,128,3,3]$ & 0 \\
ConvTranspose2d-3 & $[-1,16,9,9]$ & 51,216 \\
ConvTranspose2d-4 & $[-1,8,21,21]$ & 3,208 \\
ConvTranspose2d-5 & $[-1,1,25,25]$ & 201 \\
Upsample-6 & $[-1,1,28,28]$ & 0 \\
\hline
\multicolumn{2}{l}{Total params (trainable / non-trainable)} & 61,153 / 0 \\
\hline
\end{tabular}
\end{table}

In our denoising model, we use a convolutional neural network to implement end-to-end recovery from the measured principal component signals to the ground truth of the object. The neural network contains a fully connected layer, upsampling layers, and transposed 2D convolution layers, to progressively increase the spatial resolution while learning how to synthesize a clean $28\times 28$ image. Combining the neural network with the PCA, the overall information processing from the image to the output of the decoder neural network may be regarded as an autoencoder neural network, which is standard in image denoising.
We use the PyTorch package to implement the neural network. We summarize the decoder neural network architecture in Table~\ref{tab:nn_arch_kpca50_decoder}.

\subsubsection{Experiment results}
We present the image reconstruction results in Fig.~\ref{fig:PCA_recover_vs_m_n=1}. From top to bottom, we plot the truth, the classical benchmark with vacuum noise, and the quantum case with squeezed vacuum noise at $G=10$.
From left to right, we plot 10 examples of digit 0-9. We observe that the denoising completely fails in the classical case, while the quantum-enhanced denoised images preserve most of the information about the digits, except for extreme cases, e.g. digit $5$ here.
Note that the total pixel number here is $d=28\times 28=784$, here the quantum advantage is achieved at the cost of the quantum-squeezed source number $k\simeq 50$, significantly reduced from the pixel-wise cost $d=784$.


\section{Conclusion}
\label{sec:conclusion}
We introduced a quantum-enhanced compressive imaging framework that mitigates the scalability challenge of pixel-wise squeezing through a quantum--classical co-design. By probing only a reduced set of $k$ principal components in the $d$-dimensional image space, the required numbers of squeezed sources and homodyne detectors are dramatically reduced from $d$ to $k$. In particular, for MNIST with pixel number $d=784$, clear quantum advantages persist with only $k=15$ principal components for image classification and $k=50$ for image denoising reconstruction under a near-term feasible squeezing $G=10$. For CIFAR-10 with $d=3072$, using merely $k=50$ principal components substantially enhances the classification accuracy, 45\% for quantum $G=10$ versus 28\% for classical $G=1$. 

Instead of treating sensing and reconstruction as decoupled stages, our method jointly optimizes the quantum hardware layer (the preparation of squeezed-state sources, the PCA-guided beamsplitters for engineering the entanglement structure, and quadrature-selective homodyne readout) with the classical software layer (the downstream learning-based inference). This coupling allows image statistics learned in classical software to directly guide where limited quantum resources are deployed in hardware.

Overall, our results show that quantum squeezed states drastically enhance the image information processing, and the quantum-classical co-design is a crucial design principle for building resource-efficient scalable quantum-enhanced imaging systems.

\begin{acknowledgements}
The project is supported by DOE ARPA-E Grant No. DE-AR0002067 and Halliburton Technology Partners LLC.
\end{acknowledgements}

\newpage

\bibliographystyle{splncs}
\bibliography{references}

\end{document}